\documentclass{article}  
\usepackage{bigsky2007}
\usepackage{graphicx}
\frompage{000} \topage{000}                                              
\def\rightmark{Electro-magnetic radiation at PHENIX}
\def\leftmark{T. Sakaguchi et al.}
 
\title{Measurement of electro-magnetic radiation at PHENIX} 
\authors{
{Takao Sakaguchi for the PHENIX Collaboration
}\\[2.812mm]
{\normalsize
\hspace*{-8pt}$^1$ Brookhaven National Laboratory, Physics Department, \\
Upton, NY 11973-5000, U.S.A.\\[0.2ex] 
}}
 
\abstract{Recent results on direct photons and dileptons from the PHENIX
xperiment opened up a possibility of landscaping electro-magnetic radiation
over various kinetic energies in heavy ion collisions. A detail discussion
is given based on a review of the results.}

 
\begin{document}
 
\maketitle
\setcounter{page}{1}

\section{Introduction}\label{intro}
Electro-magnetic radiation is an excellent probe for extracting
thermodynamical information. They are emitted from all the stages of
collisions, and don't interact strongly with medium once produced.
Therefore, many heavy ion experiments have explored the
radiation.  As stated in several literatures~\cite{ref1,ref2},
the electro-magnetic radiation stands for a direct radiation from
the matter produced or a prompt reaction such as initial hard scattering.
In this sense, $\pi^0$, $\eta$, $\rho$, $\omega$ or $\phi$ mesons
decaying into photons or leptons are not defined as electro-magnetic
radiations.

The electro-magnetic radiation is primarily produced through
a Compton scattering of quarks and gluons ($qg\rightarrow q \gamma$)
and an annihilation of quarks and anti-quarks
($q\overline{q} \rightarrow g \gamma$) as leading order processes,
and the next leading
process is dominated by bremsstrahlung (fragment) ($qg \rightarrow qg\gamma$).
There is also a prediction of a jet-photon conversion process,
which occurs if a QGP is formed, by a secondary interaction of
a hard scattered parton with thermal partons in the medium~\cite{ref3,ref4}.

A calculation predicts that a photon contribution from the QGP state is
predominant in the $p_T$ range of 1$<p_T<$3\,GeV/$c$~\cite{ref5}.
The signal is usually seen after subtracting photons from known hadronic
sources. The typical signal to background ratio is $\sim$10\,\%.
For $p_T>$3\,GeV/$c$, the signal is
dominated by a contribution from initial hard scattering, and $p_T<$1\,GeV,
the signal is from hadron gas through processes of
$\pi\pi(\rho) \rightarrow \gamma \rho(\pi)$, 
$\pi K^* \rightarrow  K \gamma$ and etc..

One of the big successes by now in electro-magnetic radiation measurement is
the observation of high $p_T$ direct photons that are produced in initial
hard scattering~\cite{ref2}. The high $p_T$ hadron suppression found at RHIC
is interpreted as a consequence of an energy loss of hard-scattered partons
in the hot and dense medium. It was strongly supported by the fact that
the high $p_T$ direct photons are not suppressed and well described by
a NLO pQCD calculation.

\begin{figure}[htb]
\centering
\leavevmode\epsfxsize=12cm
\epsfbox{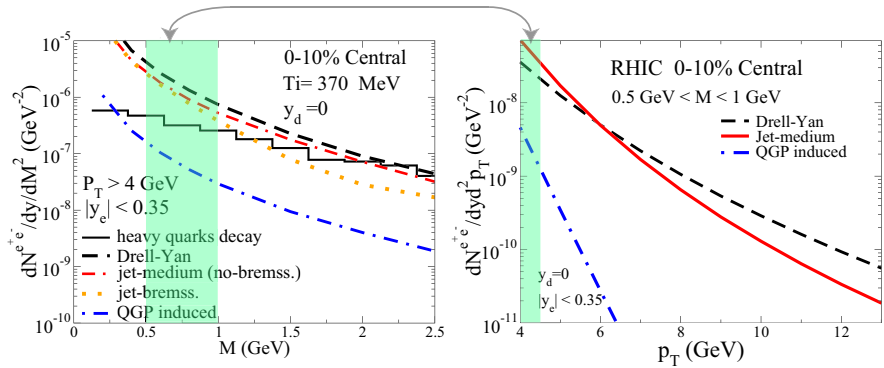}
\vspace*{-.3cm}
\caption{Prediction of dilepton yields at high mass and low $p_T$, and low mass and high $p_T$~\cite{ref6}. The shaded bands show similar kinetic regions, and one can disentangle contributions by comparing the yields.}
\vspace*{-.3cm}
\label{fig4_1}
\end{figure}

Photons are converted into virtual photons with a certain probability via
internal conversion process
(e.g. $qg \rightarrow q\gamma \rightarrow q\gamma^* \rightarrow q l^+l^-$).
This fact opened up various approaches of looking at "photons" over a broad
range of energies in a mid-rapidity; for low energy "photons"($E<$1\,GeV),
photons can be measured via low mass and low $p_T$ dileptons
(virtual photons) that decay into electrons. High energy photons ($E>$5\,GeV)
can be measured as themselves with an electro magnetic calorimeter.
In the intermediate energy region (1$<E<$5\,GeV), both dileptons and real
photons can be measured, and helps disentangling various contributions.
The idea is illustrated in Fig.~\ref{fig4_1}.

In this paper, electro-magnetic radiation is landscaped by reviewing
the latest results on direct photon and dileptons from the PHENIX experiment.

\section{Dilepton (di-electron) measurement}\label{techno}  
\begin{figure}[htb]
\centering
\begin{minipage}{60mm}
\leavevmode\epsfysize=6.0cm
\epsfbox{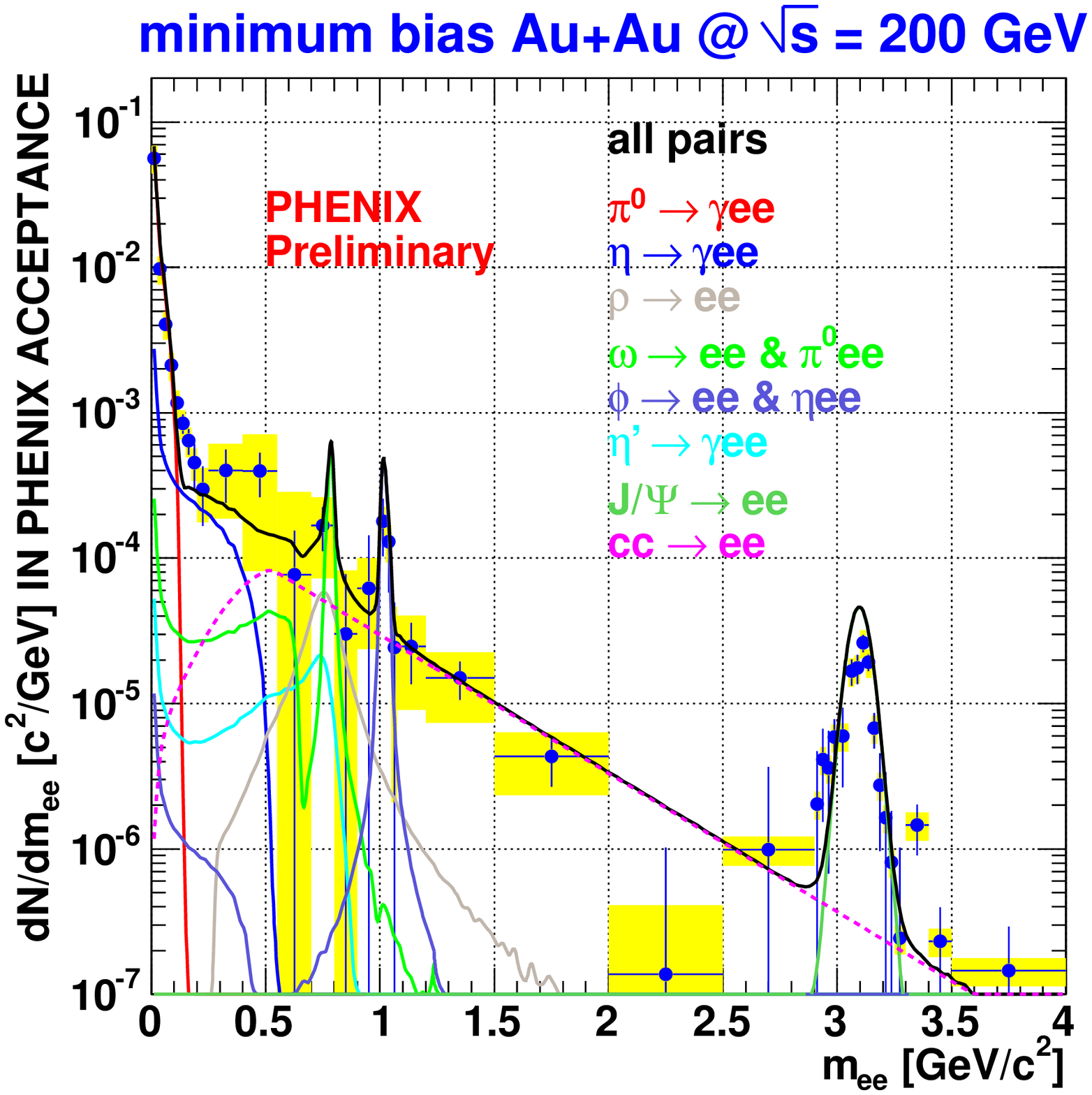}
\vspace*{-.5cm}
\vspace{-3mm}
\end{minipage}
\hspace{10mm}
\begin{minipage}{50mm}
\leavevmode\epsfxsize=4.5cm
\epsfbox{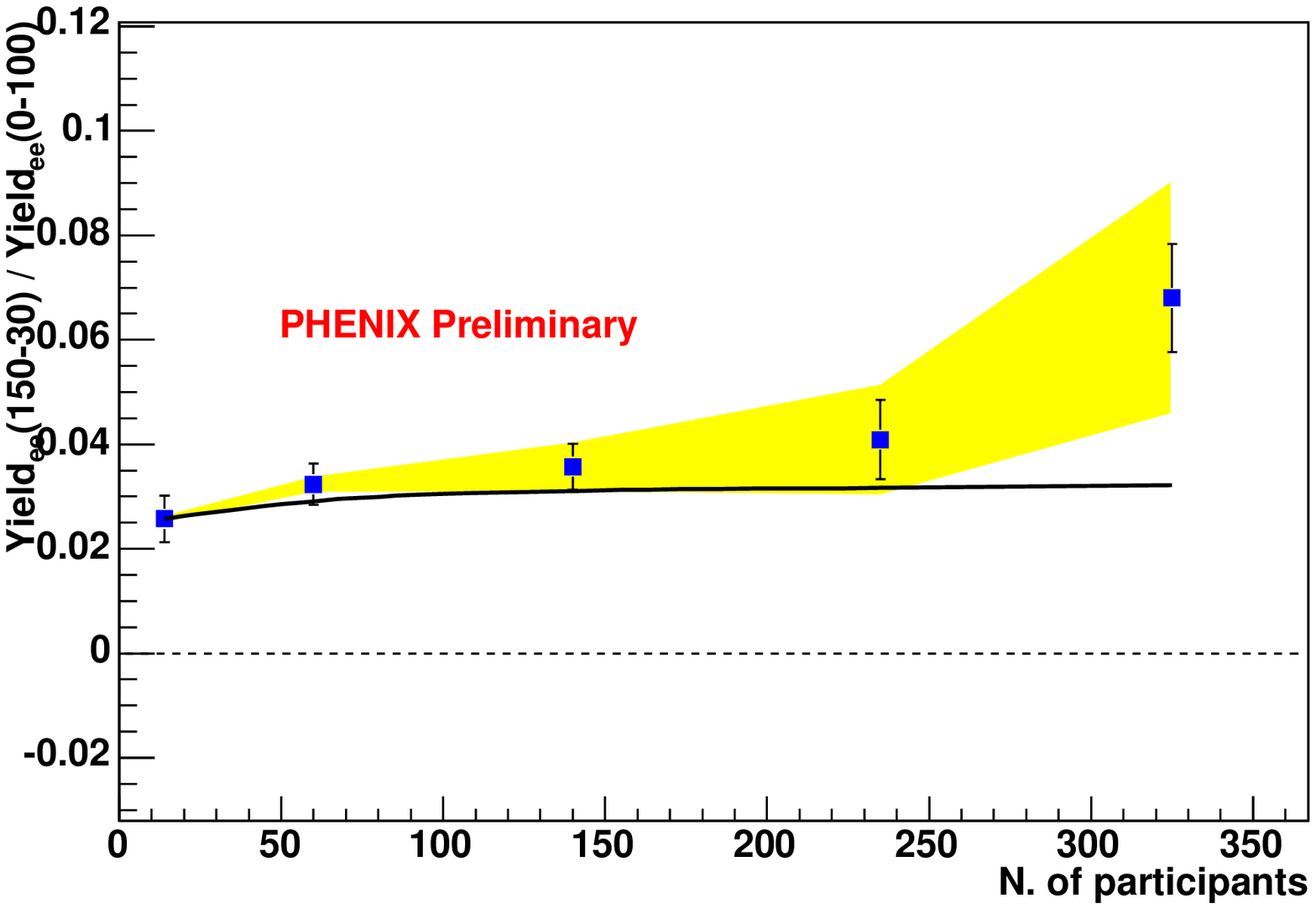}
\leavevmode\epsfxsize=4.5cm
\epsfbox{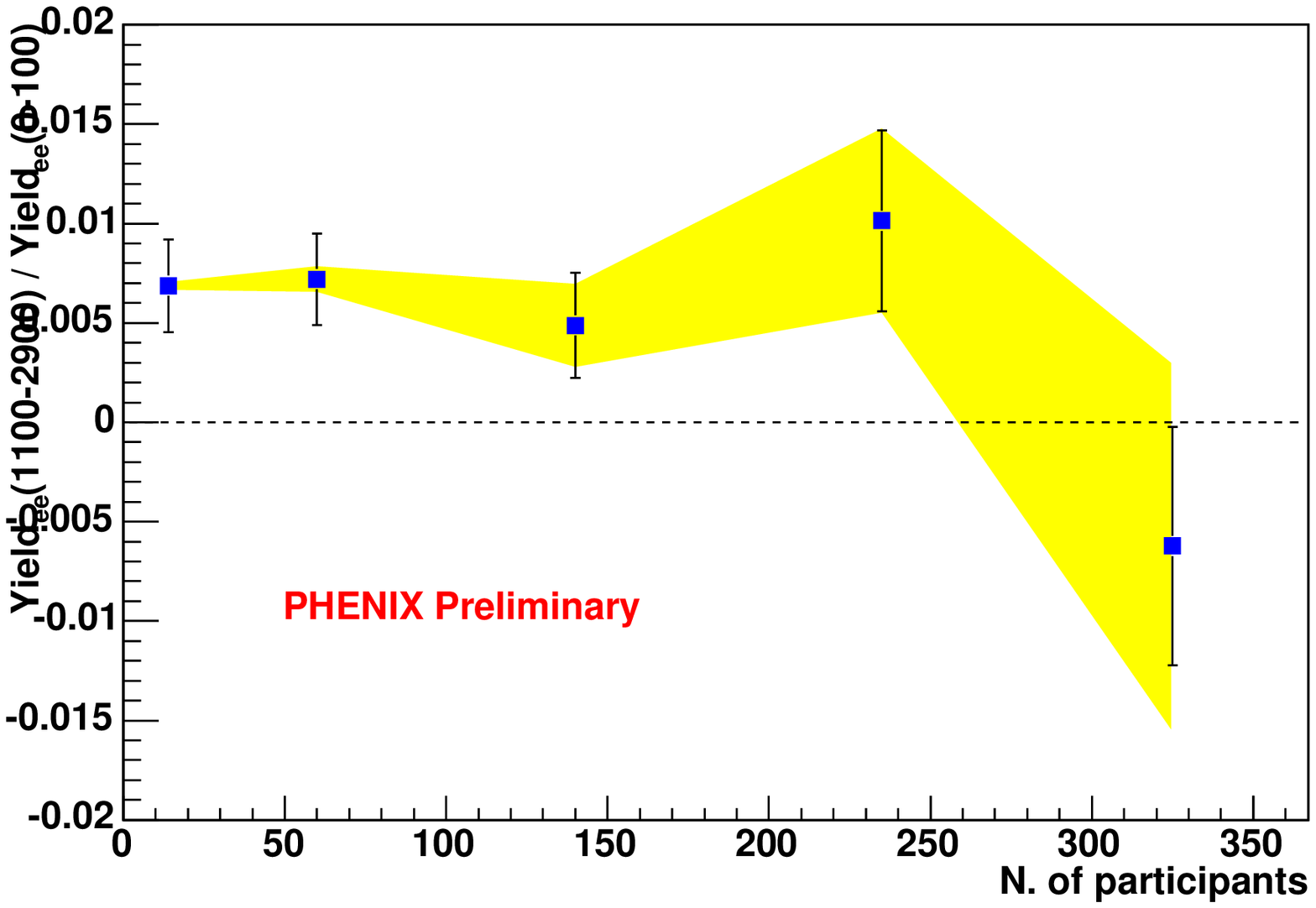}
\end{minipage}
\vspace*{-.3cm}
\caption[]{(a) Invariant mass distribution of dileptons with cocktail calculation from hadron decays (left). Ratios of yields in (b) 150-300\,MeV/$c^2$ (right top) and (c) 1.1-2.9\,GeV/$c^2$ (right bottom) to that in 0-100\,MeV/$c^2$, as a function of centrality.}
\vspace*{-.3cm}
\label{figmass}
\end{figure}

Dileptons (di-electrons) have been measured at the PHENIX experiment
using the high statistics Au+Au data in Year-4~\cite{ref12,ref13}.
Electrons are tracked by a drift chamber with an excellent
momentum resolution. A Cherenkov counter in the PHENIX experiment
that has a hadron rejection power on $10^4$ for a single track
separates electrons from $\pi^{+/-}$ well up to 4.9\,GeV/$c$.
The Fig.~\ref{figmass}(a) shows the dilepton mass spectra for $\sim$700\,M
minimum bias Au+Au events.
The $p_T$ cut of 0.3\,GeV/c is applied for single electrons. The ratios
of several mass ranges are shown in Figs.~\ref{figmass}(b) and (c).
The mass region of 0-100\,MeV/$c^2$ represents mainly a contribution from
$\pi^0$'s. In Fig.~\ref{figmass}(b), the ratio of the yields
in 150-300\,MeV/$c^2$ to 0-100\,MeV/$c^2$ is shown with the one from known
hadron decay contribution as a line. Although the systematic error is large,
there is an excess in most central collisions. This mass region corresponds
to the kinematic region where hadron-gas interaction plays a main role.

\begin{figure}[htb]
\centering
\vspace*{-.2cm}
\leavevmode\epsfxsize=8.0cm
\epsfbox{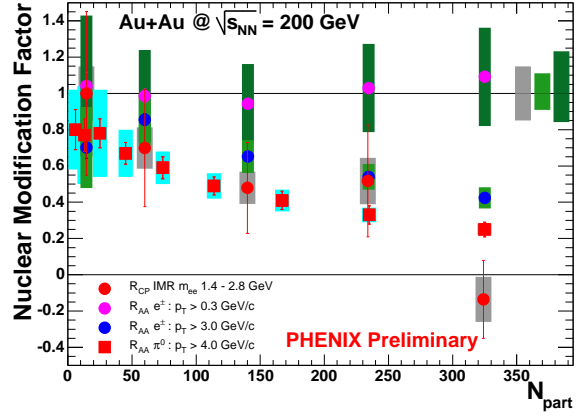}
\vspace*{-.3cm}
\caption[]{Nuclear Modification factor ($R_{AA}$) for intermediate mass
region (1.4-2.8\,GeV/$c^2$) compared with those for single electrons
($p_T>0.3$GeV\,$c$ and $p_T>2.0$GeV\,$c$)  and $\pi^0$s. Note that $R_{cp}$
(central to peripheral yield ratio) is plotted for dileptons.}
\vspace*{-.3cm}
\label{figDilepRAA}
\end{figure}

In Fig.~\ref{figmass}(c), the ratio of yields in 1.1-2.9\,GeV/$c^2$ to
0-100\,MeV/$c^2$ is shown. The ratio stays a constant up to mid-central and
drops in the most central collisions. In order to investigate the source of
contributions in the mass region, a nuclear modification factor ($R_{AA}$)
for the yield in the mass region was calculated and compared with those
for single electrons and $\pi^0$'s (Fig.~\ref{figDilepRAA}).
For dileptons, the $R_{cp}$ (central to peripheral yield ratio) is plotted
instead of $R_{AA}$ because there is no reference data from p+p collisions.
The result shows that the yield follows the suppression pattern of single
electrons and $\pi^0$'s that mainly come from semi-leptonic decay of charm
quarks and jets, respectively. The suppression of intermediate mass
dileptons could attribute to an energy loss of charm quarks, but may
also be related to alteration of an opening angle of two electrons
coming from back-to-back $c\overline{c}$ pairs.
The thermal radiation is also expected to exist in this region, but
is not clearly seen with current errors.

\section{Direct photon measurement}
\begin{figure}[htb]
\begin{minipage}{60mm}
\leavevmode\epsfxsize=6.0cm
\epsfbox{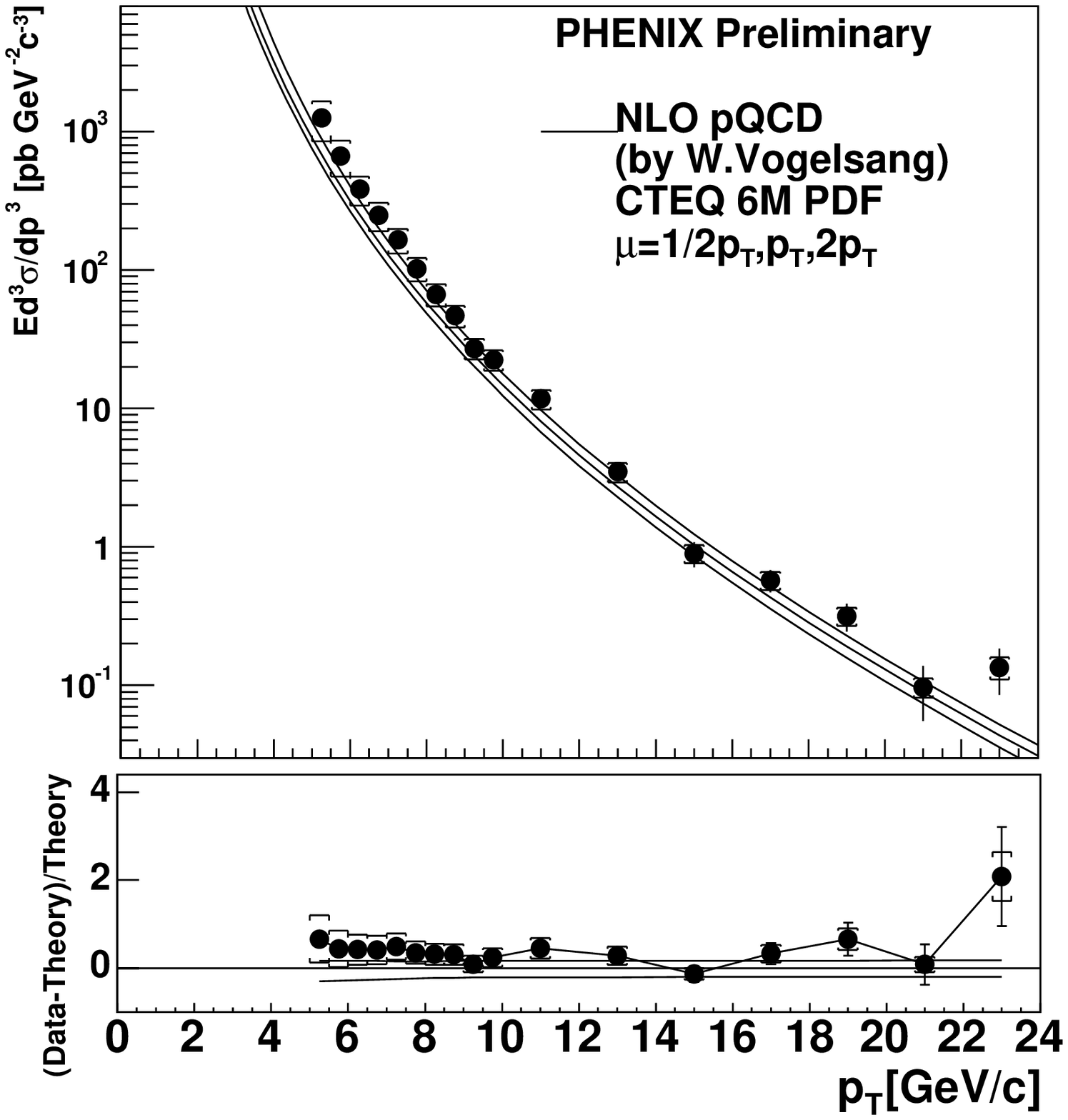}
\end{minipage}
\hspace{5mm}
\begin{minipage}{65mm}
\leavevmode\epsfxsize=6.0cm
\epsfbox{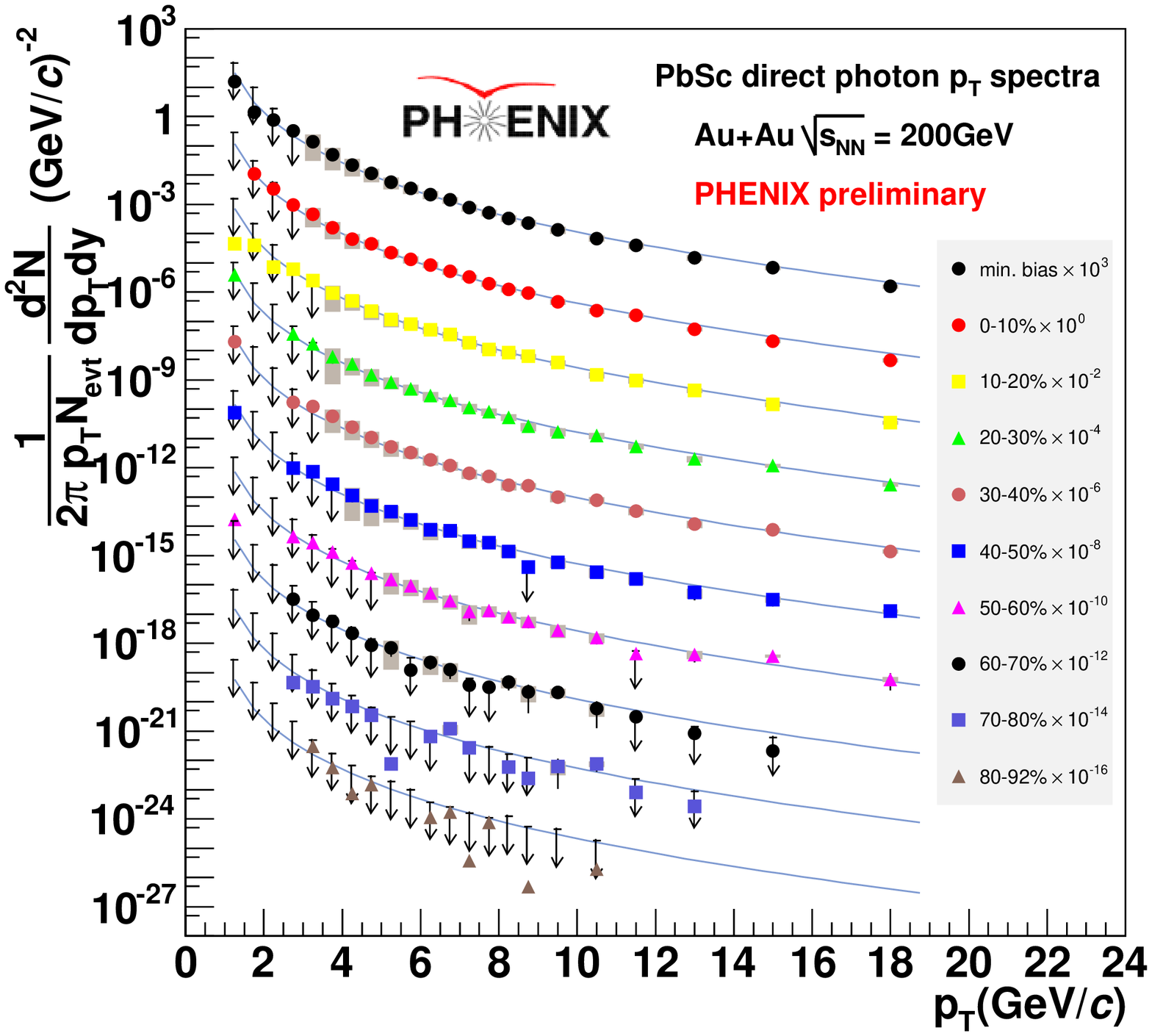}
\end{minipage}
\vspace*{-.3cm}
\caption[]{(a) Direct photon results in Au+Au collisions at $\sqrt{s_{NN}}$=200\,GeV (left) and (b) in p+p collisions at $\sqrt{s}$=200\,GeV (right).}
\vspace*{-.3cm}
\label{figphotonspectra}
\end{figure}

\begin{figure}[htb]
\begin{minipage}{60mm}
\leavevmode\epsfxsize=6.0cm
\epsfbox{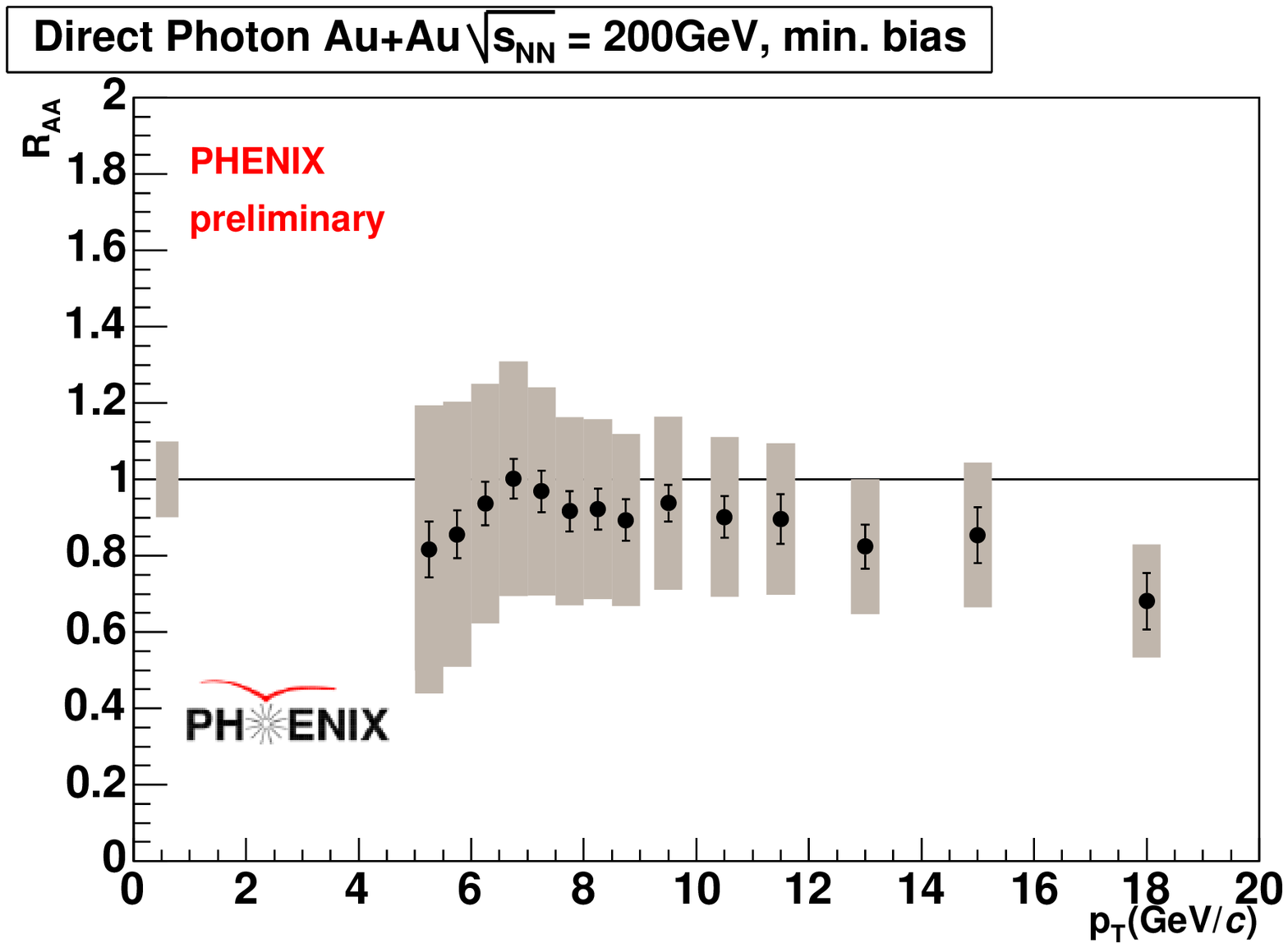}
\vspace*{-.5cm}
\vspace{-3mm}
\end{minipage}
\hspace{5mm}
\begin{minipage}{60mm}
\leavevmode\epsfxsize=6.0cm
\epsfbox{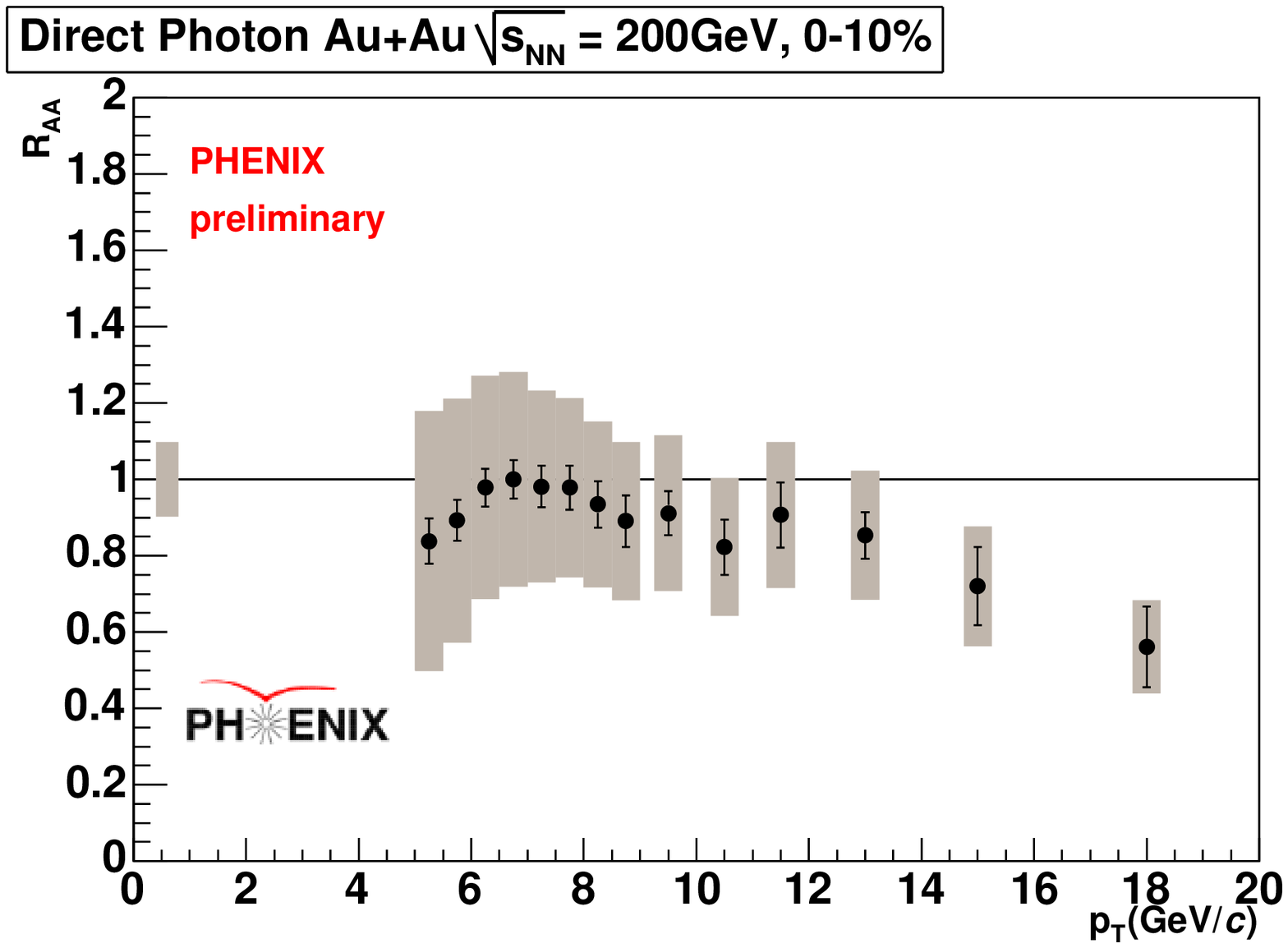}
\vspace*{-.5cm}
\vspace{-3mm}
\end{minipage}
\vspace*{-.3cm}
\caption[]{Direct photon $R_{AA}$ in (a) minimum bias (left) and (b) 0-10\,\% (right) Au+Au collisions at $\sqrt{s_{NN}}$=200\,GeV.}
\vspace*{-.3cm}
\label{figRAAphotons}
\end{figure}

Direct photons are measured in Au+Au collisions at
$\sqrt{s_{NN}}$=200\,GeV, and p+p collisions at $\sqrt{s}$=200\,GeV
as shown in Figs.~\ref{figphotonspectra}~\cite{ref14}. The direct photons
in p+p collisions are measured up to 25\,GeV/$c$, and can be used as a
reference for quantifying a medium effect in Au+Au collisions. The data is
compared with a NLO pQCD calculation~\cite{ref9}, and found that it is
well described by the calculation within $\sim$40\,\% down to 5\,GeV/$c$.
Since the $p_T$ binning is different between Au+Au and p+p results,
the p+p data is fitted with a power-law function to interpolate the
$p_T$ points of Au+Au data. The fit describes the data very well
within $\sim$5\,\%.
Fig~\ref{figRAAphotons} shows the $R_{AA}$ of direct photons in Au+Au
collisions.
In Year-2 data, we were not able to reach above 12\,GeV/$c$, where the
$R_{AA}$ was consistent with unity, and thus concluded that direct photons
are unmodified by the medium. The latest data shows a trend of decreasing
at high $p_T$ ($p_T>$14\,GeV/$c$).

\begin{figure}[htb]
\centering
\begin{minipage}{6.0cm}
\leavevmode\epsfxsize=5.5cm
\epsfbox{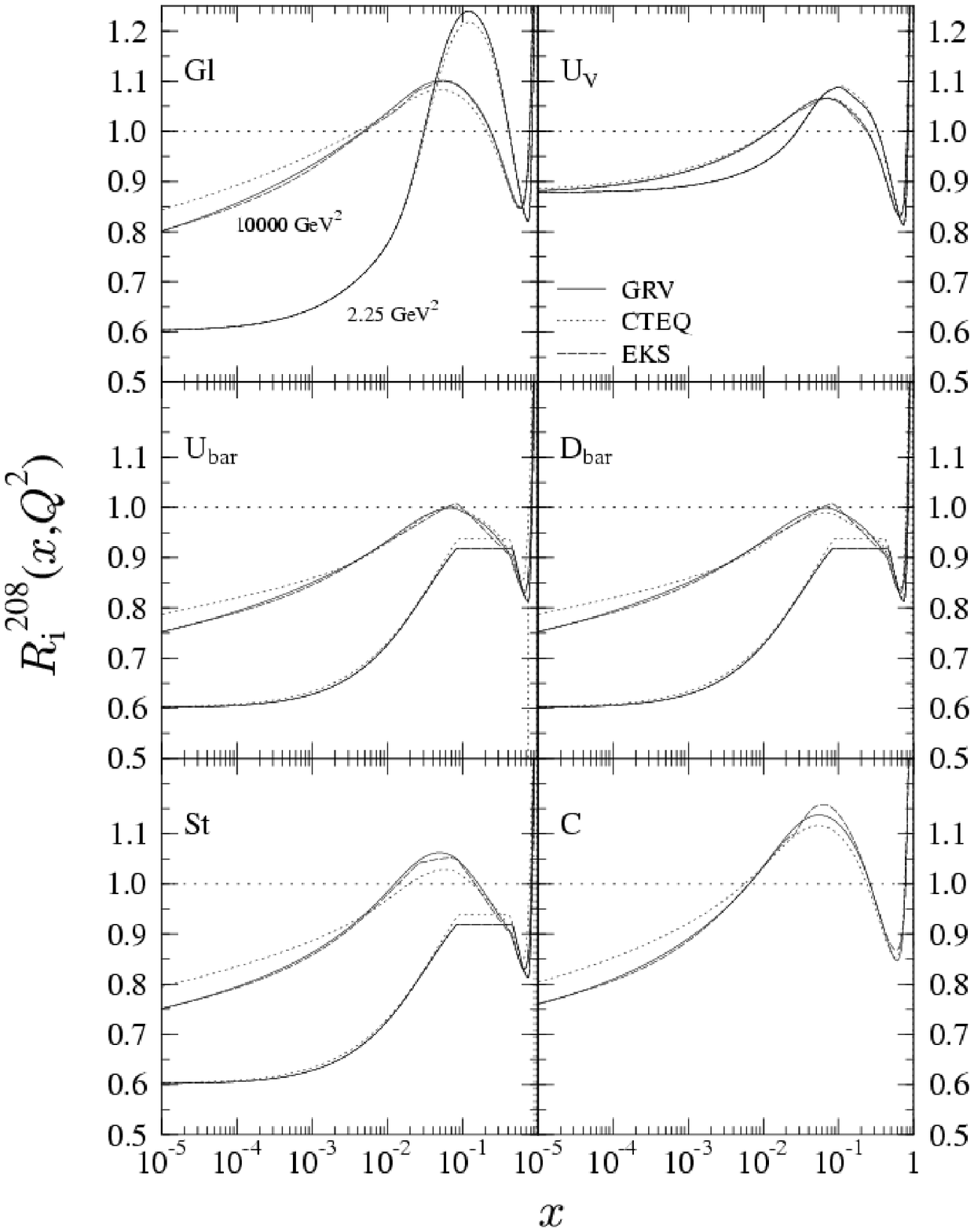}
\end{minipage}
\hspace{3mm}
\begin{minipage}{60mm}
\leavevmode\epsfxsize=5.5cm
\epsfbox{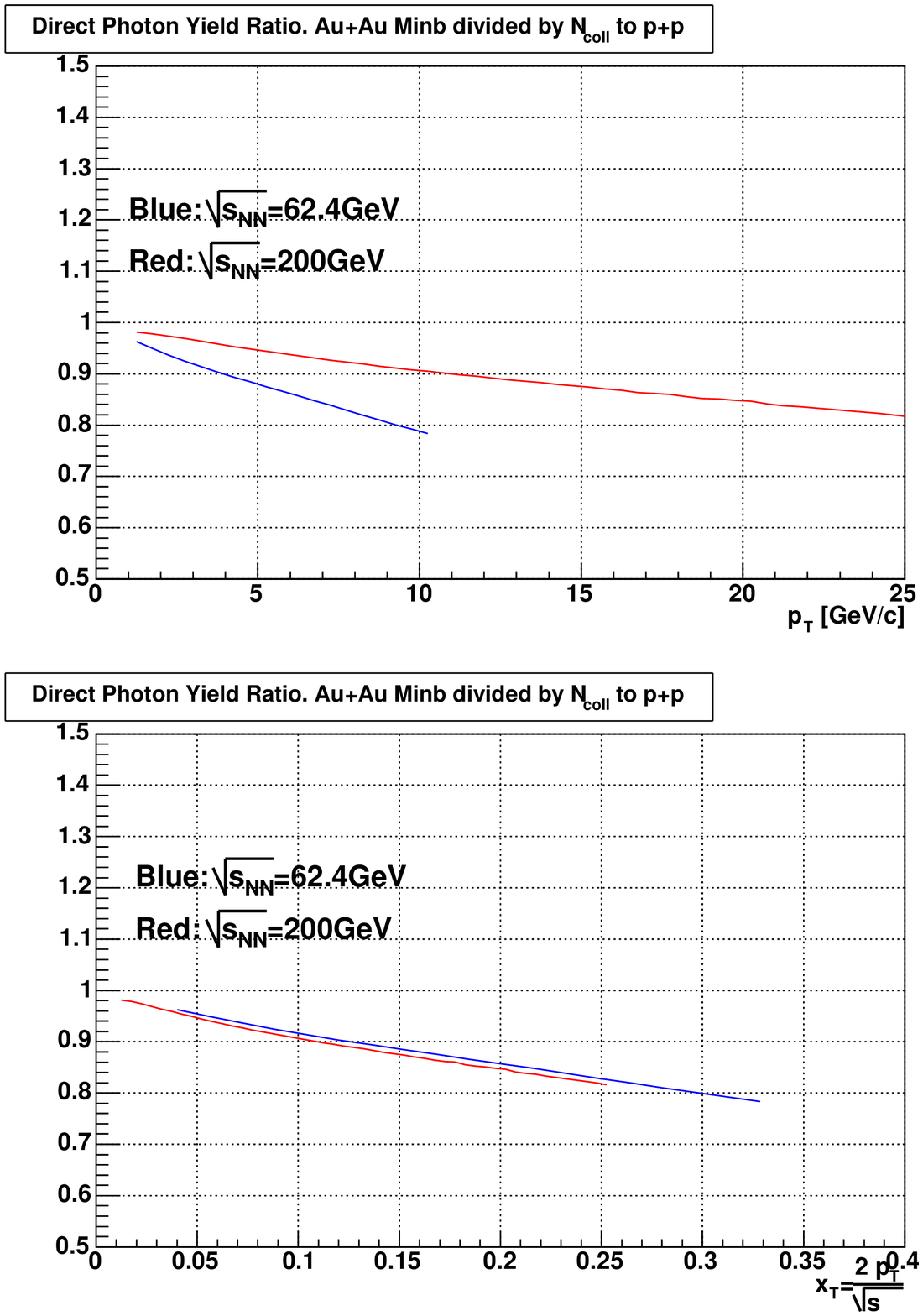}
\end{minipage}
\vspace*{-.3cm}
\caption[]{(a) Parton distribution functions and (b) expected isospin effect in Au+Au collosions calculated from n+p, p+p and n+n direct photon cross-sections.}
\vspace*{-.3cm}
\label{figStuct}
\end{figure}

There are several implications of the data such as suppression of fragment
photons (~30\,\% of total NLO pQCD photons at 14\,GeV/$c$, and decreases
as $p_T$ increases.) due to an energy loss of quarks, or an isospin
effect~\cite{ref7}. Here, simple models are proposed to understand the
result~\cite{ref16}.
The major contribution to direct photon production at the $p_T$ range of
the interest is from Compton scattering process ($qg \rightarrow q \gamma$),
therefore, we can assume that the yield is naively described as:
\[ Yield (x_T, Q^2) = F_{2p}(x_T) \times g_p(x_T) \times \sigma^{dir.\gamma}(x_T, Q^2) \]
where $F_{2p}$ is the quark parton distribution function (PDF), and $g_{p}$
is the gluon PDF. The $R_{AA}$ can be written as:
\[ R_{AA} = \frac{d^2 {\sigma_{\gamma}}^{AA}/d{p_T}^2 dy}{AA d^2 {\sigma_{\gamma}}^{pp}/d{p_T}^2 dy} \approx \left(\frac{F_{2A}(x_T)}{A F_{2p}(x_T)} \times \frac{g_{A}(x_T)}{A g_{p}(x_T)} \right) \]
The PDFs are shown in Fig.~\ref{figStuct}(a)\cite{ref8}. 
The decrease of the yield in Au+Au starts at $\sim$12\,GeV/$c$ and
drop by $\sim$30\,\% at 18\,GeV/$c$, which corresponds to x = 0.12 to 0.18.
Just from the parton distribution function, it seems that the significant
drop of $R_{AA}$ at high $p_T$ is not well explained. The structure function
can be measured in a future high statistics d+Au collisions.

The isospin effect is an effect caused from the difference of the quark
charge contents in neutrons and protons. The photon production cross-section
is proportional to $\alpha \alpha_s \Sigma e_q^2$, therefore the yield of
photons will be different between n+p, p+p and n+n collisions~\cite{ref9}.
A gold ion consists of 79 protons and 118 neutrons. We can calculate
the hard scattering cross-section for minimum bias Au+Au collisions by
weighting those for n+p, p+p and n+n as follows:
\[\frac{\sigma_{AA}}{<N_{coll}>} = \frac{1}{A^2} \times (Z^2 \sigma_{pp} + 2Z(A-Z) \sigma_{pn} + (A-Z)^2 \sigma_{nn}) \]
The $R_{AA}$ expected from isospin effect can be calculated as:
\[R_{AA} = \frac{\sigma_{AA}}{<N_{coll}>\sigma_{pp}} \]
Fig.~\ref{figStuct}(b) shows the $R_{AA}$ calculated in this way.
The calculation at $\sqrt{s_{NN}}$=200GeV is shown in red. There is
$\sim$15\,\% drop at 18\,GeV/$c$ caused by the effect. If we combine
the structure function effect with the isospin effect, the data could be
explained. It also means that the direct photons may not be modified by
a medium as expected. For a reference, the one at $\sqrt{s_{NN}}$=62.4\,GeV
is also shown as blue. It is seen that the suppression is larger at the
energy because the effect scales with $x_T$ as shown
in the bottom of Fig.~\ref{figStuct}(b). The calculation suggests that
by looking at a 62.4\,GeV result, we can quantify the isospin effect
in Au+Au collisions. The analysis is ongoing.

\section{Direct photon $v_2$}
\begin{figure}[htb]
\centering
\vspace{5mm}
\begin{minipage}{60mm}
\leavevmode\epsfxsize=5.5cm
\epsfbox{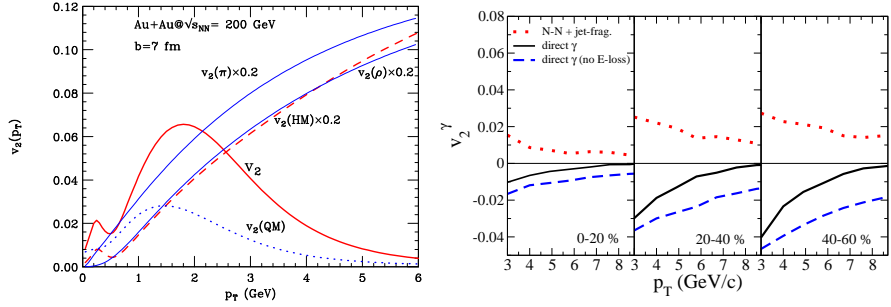}
\end{minipage}
\hspace{-5mm}
\begin{minipage}{60mm}
\leavevmode\epsfxsize=6.0cm
\epsfbox{Turbide_v2_ph_alph34_side.eps}
\end{minipage}
\caption[]{Predictions of elliptic flow of thermal (hadronic gas and partonic state) photons (left), and jet-photon conversion, Bremsstrahlung and initial hard scattering photons (right).}
\label{figTheoryFlow}
\end{figure}

The contribution of photons can be disentangled by looking at the yield
as a function of their emission angles with refer to a reaction
plane. Figs~\ref{figTheoryFlow} show predictions of elliptic flow
depending on emission sources~\cite{ref10,ref11}.
The flow of photons from hadron-gas interaction and thermal radiation
follows the collective expansion of a system, and would give a positive
$v_2$. The yield of photons produced by a Compton scattering of hard
scattered partons and medium thermal partons (jet-photon conversion)
increases as the thickness of the matter to traverse increases, and thus
gives a negative $v_2$. The bremsstrahlung photons will also increase
in out-plane, and gives a negative $v_2$. The intrinsic fragment or
bremsstrahlung photons from jets will be increased in in-plane, since
a larger energy loss of jets in out-plane will result in a lower yield of
photons originated from the jet at a given $p_T$.

\begin{figure}[htb]
\centering
\leavevmode\epsfxsize=12cm
\epsfbox{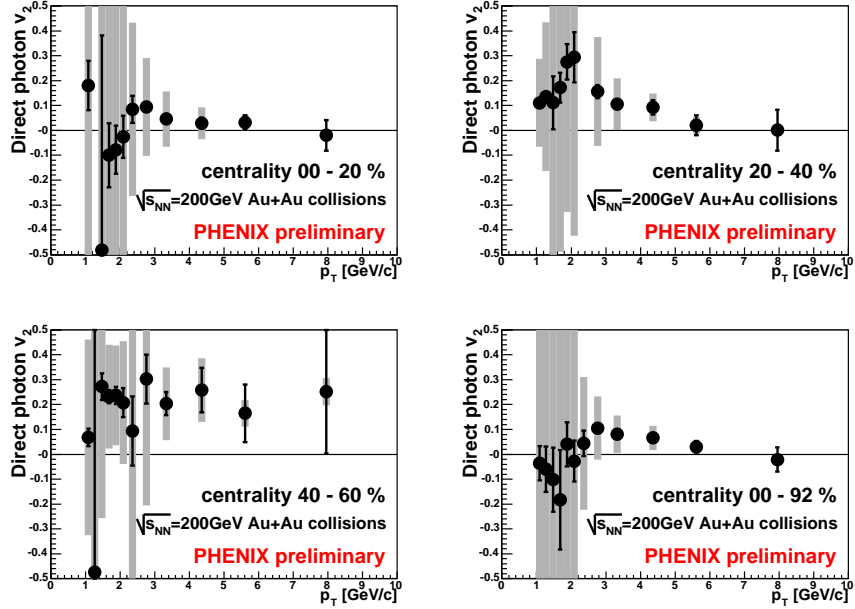}
\vspace{-5mm}
\caption[]{Elliptic flow of direct photons in Au+Au collisions at $\sqrt{s_{NN}}$=200\,GeV.}
\label{figv2PHENIX}
\end{figure}

PHENIX has measured the $v_2$ of direct photons by subtracting the
$v_2$ of hadron decay photons from that of inclusive photons as follows:
\[ {v_2}^{dir.} = \frac{{v_2}^{incl.} -{v_2}^{bkgd.}}{R-1} \]
where
\[R = \frac{(\gamma/\pi^0)_{meas}}{(\gamma/\pi^0)_{bkgd}} \]
comes from the spectral analysis~\cite{ref14}. The result is shown in
Figs.~\ref{figv2PHENIX}~\cite{ref17}.
Although the systematic error is very large, the $v_2$ of direct photons
tend to be positive in 3-6\,GeV/$c$ independent of centrality, which
is opposed to the predictions. The reduction of the systematic
errors is now ongoing to make the final conclusion.

\section{Conclusions}\label{concl}
Recent results on direct photons and dileptons from the PHENIX experiment
opened up a possibility of landscaping electro-magnetic radiation over
various kinetic energies in heavy ion collisions. A detail discussion
is given based on a review of the results. The direct photon result
in 62.4\,GeV Au+Au collisions will disentangle the effect
involved at high $p_T$.

\vfill\eject
\end{document}